\documentclass[12pt]{article}

\usepackage{amsmath}
\usepackage{hyperref}
\usepackage[T1]{fontenc}

\ifx\pdfoutput\undefined
    \usepackage[dvips]{graphicx}
\else
    \usepackage[pdftex]{graphicx}
\fi

\usepackage{amsmath}
\usepackage{amsthm}
\usepackage{amssymb}

\newcommand{\Bra}[1]{\left<#1\right|}
\newcommand{\ket}[1]{\big|#1\big>}
\newcommand{\Ket}[1]{\left|#1\right>}

\newcommand{\set}[1]{\{#1\}}
\newcommand{\Set}[1]{\left\{#1\right\}}

\newcommand{\abs}[1]{|#1|}
\newcommand{\Abs}[1]{\left|#1\right|}

\newcommand{\intervalcc}[2]{[#1,\, #2]}
\newcommand{\intervalco}[2]{[#1,\, #2)}

\newcommand{\intervaloo}[2]{(#1,\, #2)}

\newcommand{\ud}{\mathrm{d}}

\newtheorem{defi}{Definition}
\newtheorem{theo}[defi]{Theorem}
\newtheorem{lemm}[defi]{Lemma}

\usepackage[matrix,frame,arrow]{xy}
\usepackage{amsmath}
\newcommand{\qw}[1][-1]{\ar @{-} [0,#1]}
\newcommand{\qwdot}[1][-1]{\ar @{.} [0,#1]}
\newcommand{\qwx}[1][-1]{\ar @{-} [#1,0]}

\newcommand{\gate}[1]{*{\xy *+<.6em>{#1};p\save+LU;+RU **\dir{-}\restore\save+RU;+RD **\dir{-}\restore\save+RD;+LD **\dir{-}\restore\POS+LD;+LU **\dir{-}\endxy} \qw}

\newcommand{\control}{*-=-{\bullet}}

\newcommand{\ctrl}[1]{\control \qwx[#1] \qw}

\newcommand{\multigate}[2]{*+<1em,.9em>{\hphantom{#2}} \qw \POS[0,0].[#1,0];p !C *{#2},p \save+LU;+RU **\dir{-}\restore\save+RU;+RD **\dir{-}\restore\save+RD;+LD **\dir{-}\restore\save+LD;+LU **\dir{-}\restore}
\newcommand{\ghost}[1]{*+<1em,.9em>{\hphantom{#1}} \qw}

\newcommand{\rstick}[1]{*!L!<-.5em,0em>=<0em>{#1}}
\newcommand{\lstick}[1]{*!R!<.5em,0em>=<0em>{#1}}

\newcommand{\Qcircuit}{\xymatrix @*=<0em>}

\newcommand{\eps}{\varepsilon}

\begin{document}

\title{A Lower Bound for the Sturm-Liouville Eigenvalue Problem on a Quantum Computer}

\date{\today}

\author{Arvid J. Bessen\footnote{bessen@cs.columbia.edu}\\
Department of Computer Science, Columbia University}

\maketitle

\begin{abstract}
  We study the complexity of approximating the smallest eigenvalue of a univariate Sturm-Liouville problem on a quantum computer.
  This general problem includes the special case of solving a one-dimensional Schr{\"o}dinger equation with a given potential for the ground state energy.

  The Sturm-Liouville problem depends on a function~$q$, which, in the case of the Schr{\"o}dinger equation, can be identified with the potential function $V$.
  Recently Papageorgiou and Wo{\'z}\-nia\-kows\-ki proved that quantum computers achieve an exponential reduction in the number of queries over the number needed in the classical worst-case and randomized settings for smooth functions~$q$.
  Their method uses the (discretized) unitary propagator and arbitrary powers of it as a query (``power queries'').
  They showed that the Sturm-Liouville equation can be solved with $\mathcal{O}(\log (1/\eps))$ power queries, while the number of queries in the worst-case and randomized settings on a classical computer is polynomial in $1/\eps$.
  This proves that a quantum computer with power queries achieves an exponential reduction in the number of queries compared to a classical computer.

  In this paper we show that the number of queries in Papageorgiou's and Wo{\'z}\-nia\-kows\-ki's algorithm is asymptotically optimal.
  In particular we prove a matching lower bound of $\Omega(\log (1/\eps))$ power queries, therefore showing that $\Theta(\log (1/\eps))$ power queries are sufficient and necessary.
  Our proof is based on a frequency analysis technique, which examines the probability distribution of the final state of a quantum algorithm and the dependence of its Fourier transform on the input.
\end{abstract}

\section{Introduction}

This paper deals with the solution of the Sturm-Liouville problem on a quantum computer.
Quantum computers have shown great promise in solving problems as diverse as the discrete problems of searching and factoring \cite{gro-96a,sho-94} and the continuous problems including integration, path integration, and approximation \cite{nov-01,hei-01,tra-woz-01,hei-03a,hei-03b}.
The main motivation for quantum computing is its potential to solve these important problems efficiently.
Shor's algorithm achieves an exponential speedup over any known classical algorithm for factoring, but until the classical complexity of factoring is proven, the exponential speedup remains a conjecture.
The quantum algorithms for integration provide provable exponential speedups over classical \emph{worst-case} algorithms, but only polynomial speedups over classical \emph{randomized} algorithms.

Recently Papageorgiou and Wo{\'z}niakowski introduced a quantum algorithm for the Sturm-Liouville problem \cite{pap-woz-05} which uses the quantum phase estimation algorithm.
They showed that quantum algorithms with power queries\footnote{We will define power queries rigorously in Definition \ref{defi:sl-power-query}. Informally they are just an arbitrary (integer) power of a specific unitary matrix.} achieve a provable exponential reduction in the number of power queries over the number of queries needed in the classical worst-case or randomized setting.
Naturally query complexity results neglect the cost of actually implementing the queries.
At the end of this paper we will discuss this problem for power queries, but it is currently not clear under which conditions power queries are sufficiently inexpensive to implement for the Sturm-Liouville problem.

In this paper we will prove \emph{lower bounds} on the number of power queries for quantum algorithms that solve the Sturm-Liouville problem.
This can be used to show the optimality of the algorithm proposed in \cite{pap-woz-05}.
To prove lower bounds for algorithms with power queries the previously known quantum lower bound techniques, such as the ``polynomial method'' of Beals et. al \cite{bea-buh-cle-mos-wol-98, nay-wu-99} do not suffice.
Our lower bound method builds on the ``trigonometric polynomial method'' \cite{bes-04}, which is an extension of the above-mentioned polynomial method and was modified to be used with power queries in \cite{bes-04a} to prove lower bounds for the phase estimation algorithm.
Our method uses frequency analysis instead of a maximum degree argument, which is not applicable in the case of arbitrary powers.

\section{The Sturm-Liouville eigenvalue problem}

Papageorgiou and Wo{\'z}niakowski study in \cite{pap-woz-05} a simplified version of the univariate Sturm-Liou\-vil\-le problem. Consider the eigenvalue problem for the differential equation
\begin{equation}\label{eqn:sl-pap-woz}
  \begin{split}
    - u''(x) + q(x) u(x) = \lambda u(x)\\
    u(0) = u(1) = 0
  \end{split}
\end{equation}
for a given nonnegative function $q$ belonging to the class $\mathbf{Q}$ defined as
\begin{equation}\label{eqn:Q}
  \mathbf{Q}
  =
  \Big\{ q:[0,1] \to [0,1] \ : \ q\in C^2([0,1]) \text{ and } \max_{i=0,1,2} \max_{x\in [0,1]}|q^{(i)}(x)| \leq 1 \Big\} .
\end{equation}

We are looking for the smallest eigenvalue $\lambda$ such that there exists a non-zero function $u_\lambda$ that satisfies (\ref{eqn:sl-pap-woz}).
What is the minimal number of queries of $q$ that permits the determination of the smallest eigenvalue $\lambda$ in this equation with error $\eps$ and probability $3/4$ on a classical or quantum computer?

The one-dimensional time-independent Schr{\"o}dinger equation
\begin{equation}\label{eqn:schroedinger}
  - \frac{\hbar^2}{2 m} \frac{\ud^2}{\ud x^2} \Psi(x) + V(x) \Psi(x) = E \Psi(x)
\end{equation}
of a particle in a box, see \cite{mes-61}, is an instance of (\ref{eqn:sl-pap-woz}).
We are given a potential $V$ and are looking for the eigenfunctions $\Psi$ of this equation and their corresponding energies $E$.
In particular, we are interested in the ground-state and its energy, i.e., for a given potential $V$, we want to determine the eigenfunction $\Psi_0$ and its energy $E_0$, such that all other eigenfunctions $\Psi_n$ have higher energies $E_n \geq E_0$.
Since quantum systems obey equation (\ref{eqn:schroedinger}), it seems plausible that quantum computers could potentially solve the eigenvalue problem faster than a classical computer.

In the next section we define a quantum algorithm with power queries.
We especially have to tackle the question concerning the form of the input (i.e., the function $q$ in the Sturm-Liouville problem) enters the quantum algorithm.

\section{Quantum algorithms for the Sturm-Liou\-ville problem}

Let us denote the differential operator associated with the Sturm-Liouville problem for a certain $q \in \mathbf{Q}$
as $\mathbb{L}_q : C^2(\intervalcc{0}{1}) \rightarrow C^0(\intervalcc{0}{1})$, defined by
\begin{equation*}
  \mathbb{L}_q u(x) = - \frac{\ud^2}{\ud x^2} u(x) + q(x) u(x).
\end{equation*}
We discretize $\mathbb{L}_q$ by approximating the second derivative at the points $\frac{1}{n+1}$, $\frac{2}{n+1}$, $\ldots$, $\frac{n}{n+1}$ and obtain an $n \times n$ matrix $M_q$:
{\small%
  \begin{equation}\label{eqn:Mq}
    M_q = (n+1)^2
    \begin{bmatrix}
      2 & -1 & & & \\
      -1 & 2 & -1 & & \\
      & \ddots & \ddots & \ddots & \\
      & & -1 & 2 & -1 \\
      & & & -1 & 2
    \end{bmatrix}
    +
    \begin{bmatrix}
      q(\frac{1}{n+1}) \hspace{-5pt} & & & & \\
      & \hspace{-5pt} q(\frac{2}{n+1}) \hspace{-5pt} & & & \\
      & & \hspace{-5pt} \ddots \hspace{-5pt} & & \\
      & & & \hspace{-5pt} q(\frac{n-1}{n+1}) \hspace{-5pt} & \\
      & & & & \hspace{-5pt} q(\frac{n}{n+1})
    \end{bmatrix} .
  \end{equation}%
}%
The eigenvalues of $\mathbb{L}_q$ and $M_q$ are closely related.
Let us denote the smallest eigenvalue of $\mathbb{L}_q$ by $\lambda(q)$ and let us write $\lambda_1(M_q)$ for the smallest eigenvalue of $M_q$.
Then (see e.g. \cite{kel-68})
\begin{equation}\label{eqn:eigenvalue-discretization-error}
  \lambda(q) - \lambda_1(M_q) = \mathcal{O} ( n^{-2} ).
\end{equation}

The input $q \in \mathbf{Q}$
enters the quantum computer in the form of a unitary black-box transformation called a quantum query.
For the Sturm-Liouville problem we define this query to be the unitary operator $\exp( \tfrac{i}{2} M_q)$.
One can show that the smallest eigenvalue $\lambda(q)$ of the Sturm-Liouville equation satisfies $\pi^2 \leq \lambda(q) \leq \pi^2 + 1$.
To avoid ambiguity we use proper scaling, i.e., instead of $\exp(i M_q)$ we use $\exp(\tfrac{i}{2} M_q)$, which defines a unique phase $\varphi \in \intervalco{0}{1}$ by
$
  2 \pi i \varphi = \tfrac{i}{2} \lambda(q)
$.

We now define an associated quantum \emph{power query} for $\exp( \tfrac{i}{2} M_q )$.
\begin{defi}\label{defi:sl-power-query}
Let $\mathbb{L}_q$ be the differential operator for a Sturm-Liouville problem and $M_q$ its discretization at $n$ points as in (\ref{eqn:Mq}) for $q \in \mathbf{Q}$.
We define the power query $W_l^p(\exp(\tfrac{i}{2} M_q))$, where $l \in \Set{1, 2, \ldots c}$ and $p \in \mathbb{N}$, acting on $\mathbb{C}^{2^c} \otimes \mathbb{C}^n$ as
\begin{equation*}
  W_l^p(\exp(\tfrac{i}{2} M_q)) \Ket{x_1} \ldots \Ket{x_c} \Ket{\psi} = 
  \begin{cases}
    \Ket{x_1} \ldots \Ket{x_c} \exp(\tfrac{i}{2} p M_q) \Ket{\psi} & \text{for } x_l = 1 \\
    \Ket{x_1} \ldots \Ket{x_c} \Ket{\psi} & \text{otherwise}
  \end{cases}
\end{equation*}
for all $x_1, \ldots, x_c \in \Set{0,1}$ and arbitrary normalized vectors $\Ket{\psi} \in \mathbb{C}^n$ and extend this definition to all quantum states by linearity.
\end{defi}

Suppose that the $\Ket{\psi_s}$, $s=1, \ldots, n$, are the eigenvectors of $M_q$ and that $M_q \Ket{\psi_s} = \lambda_s \Ket{\psi_s}$.
Then for $\Ket{\psi} = \sum_{s=1}^n \alpha_s \Ket{\psi_s}$ and $\Ket{x} = \Ket{x_1} \ldots \Ket{x_c}$ with $x_l = 1$
\begin{equation*}
  W_l^p(\exp(\tfrac{i}{2} M_q)) \Ket{x} \Ket{\psi}
  =
  \Ket{x} \exp(\tfrac{i}{2} p M_q) \Ket{\psi}
  =
  \sum_{s=1}^n \alpha_s
  \Ket{x} e^{\tfrac{i}{2} p \lambda_s} \Ket{\psi_s} .
\end{equation*}

Quantum algorithms are products of unitary transformations.
Every quantum algorithm that approximates $\lambda(q)$ can be divided into stages that use powers of $\exp(\tfrac{i}{2} M_q)$ and therefore depend on $q$, and stages that are independent of $q$.
Let us define a quantum algorithm with power queries.

\begin{defi}\label{defi:sl-power-query-algo}
  For a Sturm-Liouville problem given by the input $q \in \mathbf{Q}$ with the solution $\lambda(q)$, we define a quantum algorithm
  \begin{equation*}
    \mathcal{A} = (\ket{\psi^{(0)}}; U_0, \ldots, U_T; l_1, p_1, \ldots, l_T, p_T; \widetilde{\lambda})
  \end{equation*}
  with $T$ power queries that solves this problem as follows.
  Let $U_0$, $U_1$, $\ldots$, $U_{T}$ be arbitrary but fixed unitary transformations and $\ket{\psi^{(0)}}$ a fixed initial state.
  Let $W_{l_j}^{p_j}(\exp(\tfrac{i}{2} M_q))$ be a power query as in Definition \ref{defi:sl-power-query}.
  A measurement of the state
  \begin{equation*}
    \ket{\psi^{(T)}(\exp(\tfrac{i}{2} M_q))}
    =
    U_{T} W_{l_{T}}^{p_{T}}(\exp(\tfrac{i}{2} M_q))
    \ldots U_1 W_{l_1}^{p_1}(\exp(\tfrac{i}{2} M_q)) U_0
    \ket{\psi^{(0)}}
  \end{equation*}
  in the standard basis yields a state $\Ket{k}$ with probability $p_{k}(q)$.
  For each $k$ compute an approximation $\widetilde{\lambda}(k) \in \mathbb{R}$ to the eigenvalue of interest $\lambda(q)$ on a classical computer.
  For every $q \in \mathbf{Q}$ the probability that an $\eps$-approximation $\widetilde{\lambda}(k)$ of $\lambda(q)$ is computed is given by
  \begin{equation}\label{eqn:prob-condition}
    \sum_{ k : | \lambda(q) - \widetilde{\lambda}(k) | < \eps} p_{k}(q) .
  \end{equation}
%
  For any algorithm $\mathcal{A}$ with $T$ power queries we define
  \begin{equation*}
    e (\mathcal{A}, T) 
    =
    \inf
    \Set{
      \eps
      \, : \,
      \eps
      \text{ chosen such that (\ref{eqn:prob-condition}) is larger than }
      \tfrac{3}{4}
      \text{ for all }
      q \in \mathbf{Q}
    }
  \end{equation*}
  as the worst-case quantum error of $\mathcal{A}$.
\end{defi}
We measure in the standard basis for convenience only; a measurement in any other basis is easily achieved by modifying the operator $U_T$ accordingly.

A model like this was introduced in \cite{bea-buh-cle-mos-wol-98} for discrete inputs $q$.
It was extended to continuous functions by Heinrich in \cite{hei-01}.
Our model is an extension of this model to incorporate power queries.

\section{Upper bounds}

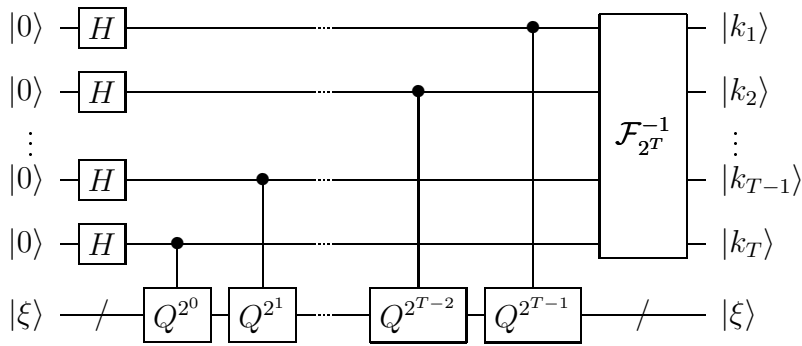
\begin{figure}[htbp]
  \begin{equation*}
    \Qcircuit @C=7pt @R=9pt {
      \lstick{\Ket{0}} & \gate{H} & \qw            & \qw            & \qw & \qwdot & \qw & \qw                & \ctrl{5}           & \multigate{4}{\mathcal{F}_{2^T}^{-1}} & \rstick{\Ket{k_{1}}} \qw \\
      \lstick{\Ket{0}} & \gate{H} & \qw            & \qw            & \qw & \qwdot & \qw & \ctrl{4}           & \qw                & \ghost{\mathcal{F}_{2^T}^{-1}}        & \rstick{\Ket{k_{2}}} \qw \\
      \lstick{\vdots\ }&          &                &                &     &        &     &                    &                    & \mathcal{F}_{2^T}^{-1}                & \rstick{\ \vdots} \\
      \lstick{\Ket{0}} & \gate{H} & \qw            & \ctrl{2}       & \qw & \qwdot & \qw & \qw                & \qw                & \ghost{\mathcal{F}_{2^T}^{-1}}        & \rstick{\Ket{k_{T-1}}} \qw \\
      \lstick{\Ket{0}} & \gate{H} & \ctrl{1}       & \qw            & \qw & \qwdot & \qw & \qw                & \qw                & \ghost{\mathcal{F}_{2^T}^{-1}}        & \rstick{\Ket{k_{T}}} \qw \\
      \lstick{\Ket{\xi}} & {/} \qw  & \gate{Q^{2^0}} & \gate{Q^{2^1}} & \qw & \qwdot & \qw & \gate{Q^{2^{T-2}}} & \gate{Q^{2^{T-1}}} & {/} \qw                               & \rstick{\Ket{\xi}} \qw \\
    }
  \end{equation*}
  \caption{The quantum phase estimation algorithm.
    $\mathcal{F}_{2^T}^{-1}$ is the inverse quantum Fourier transform on $T$ qubits.
    \label{fig:phase-est-algo}}
\end{figure}

To estimate $\lambda(q)$ on a quantum computer with power queries Papageorgiou and Wo{\'z}niakowski used the quantum phase estimation algorithm, see e.g. \cite{nie-chu-00}.
This algorithm takes a unitary transformation $Q$ with an eigenvector $\Ket{\xi}$ as input, i.e.,
$
  Q \Ket{\xi} = e^{2 \pi i \varphi} \Ket{\xi}
$.
Here $\varphi \in \intervalco{0}{1}$ is called the ``phase'' of the eigenvalue corresponding to $\Ket{\xi}$, and the phase estimation algorithm gives us an approximation $\widetilde{\varphi}$ to $\varphi$.
This algorithm has the final state
\begin{equation*}
  \Ket{\psi^{(T)}(Q)}
  =
  (\mathcal{F}_{2^T}^{-1} \otimes I) W_1^{2^{T-1}}(Q) W_2^{2^{T-2}}(Q) \ldots W_T^{2^0}(Q) (H^{\otimes T} \otimes I) \Ket{0} \Ket{\xi},
\end{equation*}
and is depicted in Figure \ref{fig:phase-est-algo}.
Suppose $Q$ is a $r$ qubit transformation.
A measurement of $\Ket{\psi^{(T)}(Q)}$ returns a state
\begin{equation*}
  \Ket{k} = \Ket{k_1} \ldots \Ket{k_T} \Ket{k_{T+1}} \ldots \Ket{k_{T+r}}.
\end{equation*}
The algorithm then uses $k$ to compute an approximation
$
\widetilde{\varphi}(k) = k_{1} 2^{-1} + k_{2} 2^{-2} + \ldots + k_T 2^{-T}
$ to $\varphi$ classically.

One can show, see e.g. \cite{nie-chu-00}, that with probability greater than $\frac{3}{4}$ the algorithm approximates $\varphi$ up to precision $\eps$ with $\mathcal{O} ( \log ( (1/\eps) ) )$ power queries.
Papageorgiou and Wo{\'z}niakowski use this algorithm to approximate the smallest eigenvalue $\lambda(q)$ of the Sturm-Liouville operator $\mathbb{L}_q$ and use the operator
$Q = \exp( \tfrac{i}{2} M_q )$
as a query.
Since the phases of $\exp( \tfrac{i}{2} M_q )$ and $\exp( \tfrac{i}{2} \mathbb{L}_q )$ are related through equation (\ref{eqn:eigenvalue-discretization-error}), we have to discretize at $n = \mathcal{O}(\eps^{-1/2})$ points.

The quantum phase estimation algorithm requires the knowledge of the eigenvector for which the phase is estimated.
For the Sturm-Liouville problem we need the eigenvector $\Ket{z_1(M_q)}$ of $M_q$ corresponding to the smallest eigenvalue $\lambda_1(M_q)$.
We can compute $\Ket{z_1(M_q)}$ through the method of Jaksch and Papageorgiou \cite{jak-pap-03}, which computes a superposition of eigenvectors $\Ket{z_j(M_q)}$ of $M_q$, with a sufficiently large $\Ket{z_1(M_q)}$ component.
For details see \cite{jak-pap-03,pap-woz-05}.

\section{Lower Bounds}

Our goal is to prove that the algorithm described in the previous section is optimal with respect to the number of power queries.
We have to prove that every quantum algorithm $\mathcal{A}$ with $T$ power queries that returns a correct answer with precision $e(\mathcal{A},T) \leq \eps$ has to use $T = \Omega (\log (1/\eps) )$ power queries.

We will show that even for a much simplified version of the problem this lower bound still holds.
Consider as input only constant functions $q(x)=q \in \intervalcc{0}{1}$.
Obviously $q \in \mathbf{Q}$.
It is easy to see that in this case the eigenfunctions which fulfill the boundary condition in (\ref{eqn:sl-pap-woz}) are
\begin{equation}\label{eqn:eigfunc-Lq}
  u_s(x) = \sin (s \pi x)
\end{equation}
for $s \in \mathbb{N}$ and that they have eigenvalues $\lambda_s = s^2 \pi^2 + q$, which means that the smallest eigenvalue $\lambda(q)$ is $\lambda(q) = \pi^2 + q$.

Similarly for the discretization $M_q$ of $\mathbb{L}_q$ with constant $q \in \intervalcc{0}{1}$ the eigenvectors are
\begin{equation}\label{eqn:eigvec-Mq}
  \Ket{u_s} = \sqrt{\tfrac{2}{n+1}} \sum_{x=1}^n \sin \big( \tfrac{s \pi x}{n+1} \big) \Ket{x}
\end{equation}
with eigenvalues $4 (n+1)^2 \sin^2 \big( \tfrac{s \pi}{2(n+1)} \big) + q$.

We want to investigate how different power queries lead to different outputs and turn to the techniques in \cite{bes-04a}.
\begin{theo}\label{theo:trig-poly-ap}
  Any quantum algorithm with power queries $W_l^p(\exp(\tfrac{i}{2} M_q))$ for $q(x) = q \in \intervalco{0}{1}$, see Definition \ref{defi:sl-power-query-algo}, that uses $c \in \mathbb{N}$ control qubits, can be written as
  \begin{equation}\label{eqn:psiTMq}
  \begin{split}
    \Ket{\psi^{(T)}(\exp(\tfrac{i}{2} M_q))}
    & = U_{T} W_{l_{T}}^{ p_{T}}(\exp(\tfrac{i}{2} M_q)) \ldots U_1 W_{l_1}^{ p_1}(\exp(\tfrac{i}{2} M_q)) U_0 \ket{\psi^{(0)}} \\
    & = \sum_{k=0}^{n 2^c - 1} S_k^{(T)} (q) \Ket{k} ,
  \end{split}
  \end{equation}
  where $U_1, \ldots, U_T$ are unitary operators and the $S_k^{(T)} (q)$ are trigonometric polynomials of the following form:
  \begin{equation}\label{eqn:sktq}
    S_k^{(T)} (q)
    = \sum_{m \in \mathcal{M}_{T}}
    \eta^{(T)}_{k,m}
    e^{ \frac{i}{2} m q} ,
  \end{equation}
  with $\mathcal{M}_T$ defined as $\mathcal{M}_0 = \Set{0}$ and
  \begin{equation}\label{eqn:M-recursive}
    \mathcal{M}_{T+1} = \Set{ m \, : \, m \in \mathcal{M}_T } \cup \Set{ m + p_{T+1} \, : \, m \in \mathcal{M}_T },
  \end{equation}
  and the coefficients $\eta^{(T)}_{k,m} \in \mathbb{C}$ do not depend on $q$ and are normalized:
  \begin{equation}\label{eqn:eta-norm}
    \sum_{k}
    \sum_{m \in \mathcal{M}_{T}}
    \abs{\eta^{(T)}_{k,m}}^2
    =
    1 .
  \end{equation}
\end{theo}
\begin{proof}
  The proof is by induction on the number of queries $T$.
  We will write the state of the algorithm after $T$ steps $\Ket{\psi^{(T)}(\exp(\tfrac{i}{2} M_q))}$ in the basis $(\Ket{k}\Ket{\psi_s})_{k,s}$, $k=0, 1, \ldots, 2^c-1$, $s=1, 2, \ldots, n$, which is split into a control part $\Ket{k}$ and an eigenvector part $\Ket{\psi_s}$.
  We will not address the ancilla qubits in our proof, but they can easily be treated (after possibly reordering the qubits) as control bits that are never used.

  For $T=0$ power queries we can write
  \begin{equation*}
    \Ket{\psi^{(0)}(\exp(\tfrac{i}{2} M_q))}
    =
    U_0 \Ket{\psi^{(0)}}
    =
    \sum_{k,s}
    \eta_{k,s,0}^{(0)}
    \Ket{k} \Ket{\psi_s}, 
  \end{equation*}
  which contains only powers $e^{\frac{i}{2} m q}$ from $m \in \mathcal{M}_0 = \Set{ 0 }$ and obviously
  \begin{equation*}
    \sum_{k,s}
    \sum_{m \in \mathcal{M}_0}
    \abs{\eta_{k,s,m}^{(0)}}^2
    =
    \sum_{k,s}
    \abs{\eta_{k,s,0}^{(0)}}^2
    = 1.
  \end{equation*}

  Let us now assume $\Ket{\psi^{(T)}(\exp(\tfrac{i}{2} M_q))}$ can be written as
  \begin{equation*}
    \ket{\psi^{(T)}(\exp(\tfrac{i}{2} M_q))}
    =
    \sum_{k,s}
    \sum_{m \in \mathcal{M}_{T}}
    \eta^{(T)}_{k,s,m}
    e^{ \frac{i}{2} m q}
    \Ket{k} \Ket{\psi_s}, 
  \end{equation*}
  with coefficients $\eta^{(T)}_{k,s,m}$ fulfilling condition (\ref{eqn:eta-norm}).
  If we apply $W_{l_{T+1}}^{p_{T+1}}(\exp(\tfrac{i}{2} M_q))$ to $\ket{\psi^{(T)}(\exp(\tfrac{i}{2} M_q))}$ we get ($k_{l_{T+1}}$ is the control bit, i.e., the $l_{T+1}$-th bit in the binary representation of $k$):
  \begin{multline}\label{eqn:W-applied}
      W_{l_{T+1}}^{p_{T+1}}(\exp(\tfrac{i}{2} M_q))
      \ket{\psi^{(T)}(\exp(\tfrac{i}{2} M_q))}
      =
      \sum_{\substack{k,s\\k_{l_{T+1}} = 0}}
      \sum_{m \in \mathcal{M}_{T}}
      \eta^{(T)}_{k,s,m}
      e^{ \frac{i}{2} m q }
      \Ket{k} \Ket{\psi_s}  \\
      + 
      \sum_{\substack{k,s\\k_{l_{T+1}} = 1}}
      \sum_{m \in \mathcal{M}_{T}}
      \eta^{(T)}_{k,s,m}
      e^{ \frac{i}{2} m q}
      \Ket{k} \exp( \tfrac{i}{2} p_{T+1} M_q ) \Ket{\psi_s} .
  \end{multline}
  We define
  $\zeta_s := e^{ \frac{i}{2} 4 (n+1)^2 \sin^2 \big( \tfrac{s \pi}{2(n+1)} \big) }$
  and proceed to analyze the second term in (\ref{eqn:W-applied}), where the control bit $k_{l_{T+1}} = 1$ and get the following
  \begin{equation*}
    \begin{split}
      &
      \sum_{m \in \mathcal{M}_{T}}
      \eta^{(T)}_{k,s,m}
      e^{ \frac{i}{2} m q}
      \Ket{k} \exp(\tfrac{i}{2} p_{T+1} M_q) \Ket{\psi_s} \\
      = &
      \sum_{m \in \mathcal{M}_{T}}
      \eta^{(T)}_{k,s,m}
      e^{ \frac{i}{2} m q}
      e^{ \frac{i}{2} p_{T+1} \big( 4 (n+1)^2 \sin^2 \big( \tfrac{s \pi}{2(n+1)} \big) + q \big) }
      \Ket{k} \Ket{\psi_s} \\
      = &
      \sum_{m \in \mathcal{M}_{T}}
      \eta^{(T)}_{k,s,m}
      \zeta_s^{p_{T+1}}
      e^{ \frac{i}{2} (m + p_{T+1}) q }
      \Ket{k} \Ket{\psi_s}
      .
    \end{split}
  \end{equation*}
  If we define $\widetilde{\eta}^{(T+1)}_{k,s,m}$ for all $m \in \mathcal{M}_{T+1}$ as
  \begin{equation*}
    \widetilde{\eta}^{(T+1)}_{k,s,m}
    :=
    \begin{cases}
    \eta^{(T)}_{k,s,m-p_{T+1}} \zeta_s^{p_{T+1}} & \text{ for } k_{l_{T+1}} = 1 \text{ and } m - p_{T+1} \in \mathcal{M}_T \\
    \eta^{(T)}_{k,s,m} & \text{ for } k_{l_{T+1}} = 0 \text{ and } m \in \mathcal{M}_{T} \\
    0 & \text{ otherwise}    
    \end{cases} ,
  \end{equation*}
  we can write
  \begin{equation*}
    W_{l_{T+1}}^{p_{T+1}}(\exp(\tfrac{i}{2} M_q))
    \ket{\psi^{(T)}(\exp(\tfrac{i}{2} M_q))}
    =
    \sum_{k,s}
    \sum_{m \in \mathcal{M}_{T+1}}
    \widetilde{\eta}^{(T+1)}_{k,s,m}
    e^{ \frac{i}{2} m q}
    \Ket{k} \Ket{\psi_s}
    .
  \end{equation*}
  We check our normalization condition (\ref{eqn:eta-norm}) for $\widetilde{\eta}^{(T+1)}_{k,s,m}$,
  \begin{equation*}
    \begin{split}
      &
      \sum_{k,s}
      \sum_{m \in \mathcal{M}_{T+1}}
      \abs{\widetilde{\eta}^{(T+1)}_{k,s,m}}^2 \\
      = &
      \sum_{\substack{k,s\\k_{l_{T+1}}=0}}
      \sum_{m \in \mathcal{M}_{T}}
      \abs{\eta^{(T)}_{k,s,m}}^2
      +
      \sum_{\substack{k,s\\k_{l_{T+1}}=1}}
      \sum_{m - p_{T+1} \in \mathcal{M}_{T}}
      \abs{\eta^{(T)}_{k,s,m - p_{T+1}} \zeta_s^{p_{T+1}}}^2 \\
      = &
      \sum_{k,s}
      \sum_{m \in \mathcal{M}_{T}}
      \abs{\eta^{(T)}_{k,s,m}}^2
      =
      1.
    \end{split}
  \end{equation*}
  
  The next step in the algorithm is to apply the unitary transformation $U_{T+1}$. 
  For $k,l = 0, \ldots, 2^c-1$ and $s,t=1, \ldots, n$
  define the coefficients
  $u_{l,t,k,s} = \Bra{l}\Bra{\psi_t} U_{T+1} \Ket{k}\Ket{\psi_s}$
  and let
  \begin{equation*}
    \eta_{l,t,m}^{(T+1)}
    :=
    \sum_{k,s}
    \widetilde{\eta}^{(T+1)}_{k,s,m}
    u_{l,t,k,s}
  \end{equation*}
  This allows us to write
  \begin{equation*}
    \begin{split}
      &
      U_{T+1} W_{l_{T+1}}^{p_{T+1}}(\exp(\tfrac{i}{2} M_q)) \ket{\psi^{(T)}(\exp(\tfrac{i}{2} M_q))} \\
      = &
      \sum_{k,s}
      \sum_{m \in \mathcal{M}_{T+1}}
      \widetilde{\eta}^{(T+1)}_{k,s,m}
      e^{ \frac{i}{2} m q}
      U_{T+1} \Ket{k} \Ket{\psi_s} \\
      = &
      \sum_{l,t}
      \sum_{m \in \mathcal{M}_{T+1}}
      \sum_{k,s}
      \widetilde{\eta}^{(T+1)}_{k,s,m}
      u_{l,t,k,s}
      e^{ \frac{i}{2} m q}
      \Ket{l} \Ket{\psi_t} \\
      = &
      \sum_{l,t}
      \sum_{m \in \mathcal{M}_{T+1}}
      \eta^{(T+1)}_{l,t,m}
      e^{ \frac{i}{2} m q}
      \Ket{l} \Ket{\psi_t} .
    \end{split}
  \end{equation*}
  It remains to check that
  \begin{equation*}
    \begin{split}
      &
      \sum_{l,t}
      \sum_{m \in \mathcal{M}_{T+1}}
      \hspace{-3pt}
      \Abs{\eta^{(T+1)}_{l,t,m}}^2
      \\
      = &
      \sum_{l,t}
      \sum_{m \in \mathcal{M}_{T+1}}
      \left[
      \sum_{k,s}
      \Big( \widetilde{\eta}^{(T+1)}_{k,s,m} \Big)^{\ast}
      \Big( u_{l,t,k,s} \Big)^{\ast}
      \right]
      \hspace{-3.5pt}
      \left[
      \sum_{k',s'}
      \widetilde{\eta}^{(T+1)}_{k',s',m}
      u_{l,t,k',s'}
      \right] \\
      = &
      \sum_{k,s,k',s'}
      \sum_{m \in \mathcal{M}_{T+1}}
      \Big( \widetilde{\eta}^{(T+1)}_{k,s,m} \Big)^{\ast}
      \left[
      \sum_{l,t}
      \Big( u_{l,t,k,s} \Big)^{\ast}
      u_{l,t,k',s'}
      \right]
      \widetilde{\eta}^{(T+1)}_{k',s',m} \\
      = &
      \sum_{k,s}
      \sum_{m \in \mathcal{M}_{T+1}}
      \Abs{ \widetilde{\eta}^{(T+1)}_{k,s,m} }^2
      =
      1 ,
    \end{split}
  \end{equation*}
  where we used that $U_{T+1}$ is unitary.
  This completes the proof.
\end{proof}

We can use Theorem \ref{theo:trig-poly-ap} to get explicit formulas for the probability of measuring a certain state.

\begin{lemm}\label{lemm:pbc}
  Let $\mathcal{A}$ be a $T$ power query quantum algorithm for the Sturm-Liouville problem with powers $p_1, \ldots, p_T$ and $c \in \mathbb{N}$ control bits as defined in Definition \ref{defi:sl-power-query-algo}.
  Let $\mathcal{B}$ be a partition of the set of all basis vectors, i.e.
  \begin{equation*}
    \bigcup_{B \in \mathcal{B}} B = \Set{ \Ket{k} \, : \, k = 0, 1, \ldots, n 2^c -1 }
    \text{ and }
    B \cap C = \emptyset
    \text{ for } B, C \in \mathcal{B},\ 
    B \neq C .
  \end{equation*}
  If the input $q \in \mathbf{Q}$ is a constant function $q(x) = q \in \intervalco{0}{1}$, the probability of measuring a state $\Ket{k}$ from $B \in \mathcal{B}$ is a trigonometric polynomial
  \begin{equation}\label{eqn:pbc}
    p_B (q)
    =
    \sum_{l \in \mathcal{L}_{T}}
    \beta^{(T)}_{B,l}
    e^{ \tfrac{i}{2} l q} ,
  \end{equation}
  with coefficients $\beta^{(T)}_{B,l} \in \mathbb{C}$ that are bounded by
  \begin{equation*}
    \sum_{B \in \mathcal{B}}
    \abs{\beta^{(T)}_{B,l}}
    \leq
    1
  \end{equation*}
  for all possible partitions $\mathcal{B}$, and the set $\mathcal{L}_T$ is given by $\mathcal{L}_0 = \Set{0}$ and
  \begin{equation}\label{eqn:L_T}
    \mathcal{L}_{T+1}
    =
    \bigcup_{ l \in \mathcal{L}_T }
    \Set{ l, l+p_{T+1}, l-p_{T+1} } .
  \end{equation}
\end{lemm}
\begin{proof}
  Consider quantum queries $\exp (\tfrac{i}{2} M_q)$ for constant functions $q(x) = q \in \intervalco{0}{1}$ in the Sturm-Liouville problem.
  From equations (\ref{eqn:psiTMq}), (\ref{eqn:sktq}) we know that the final state of every $T$ power query algorithm can be written as
  \begin{equation*}
    \Ket{\psi^{(T)}(\exp(\tfrac{i}{2} M_q))}
    =
    \sum_{k}
    \sum_{m \in \mathcal{M}_{T}}
    \eta^{(T)}_{k,m}
    e^{ \tfrac{i}{2} m q }
    \Ket{k}.
  \end{equation*}
  Let $\mathcal{B}$ be a partition of the set of all basis states $\Ket{k}$.
  Thus the probability to measure a state from the set $B \in \mathcal{B}$ is
  \begin{equation*}
    \begin{split}
      p_B (q) 
      = &
      \sum_{k \in B} \Abs{
	\sum_{m \in \mathcal{M}_{T}}
	\eta^{(T)}_{k,m}
	e^{ \tfrac{i}{2} m q }
      }^2 \\
      = &
      \sum_{k \in B} \left[ 
	\sum_{m_1 \in \mathcal{M}_{T}}
	\left( \eta^{(T)}_{k,m_1} \right)^{\ast}
	e^{- \tfrac{i}{2} m_1 q }
	\right]
      \left[
	\sum_{m_2 \in \mathcal{M}_{T}}
	\eta^{(T)}_{k,m_2}
	e^{ \tfrac{i}{2} m_2 q }
	\right] \\
      = &
      \sum_{k \in B}
      \sum_{m_1,m_2 \in \mathcal{M}_{T}}
      \left(\eta^{(T)}_{k,m_1}\right)^{\ast}
      \eta^{(T)}_{k,m_2}
      e^{ \tfrac{i}{2} (m_2-m_1) q } \\
      =: &
      \sum_{l \in \mathcal{L}_{T}}
      \beta^{(T)}_{B,l}
      e^{ \tfrac{i}{2} l q } ,
    \end{split}
  \end{equation*}
  with coefficients $\beta^{(T)}_{B,l}$ defined as
  \begin{equation}\label{eqn:beta-def}
    \beta^{(T)}_{B,l}
    :=
    \sum_{k \in B}
    \sum_{\substack{m_1, m_2 \in \mathcal{M}_{T}\\m_2 - m_1 = l}}
    \left( \eta^{(T)}_{k,m_1} \right)^{\ast}
    \eta^{(T)}_{k,m_2} ,
  \end{equation}
  and the set $\mathcal{L}_T$ is given by
  \begin{equation}\label{eqn:L_Tv2}
    \mathcal{L}_T = \Set{m_1 - m_2 \, : \, m_1, m_2 \in \mathcal{M}_T }.
  \end{equation}
  For any partition $\mathcal{B}$ we can now bound the $\beta^{(T)}_{B,l}$ as follows
  \begin{equation*}
    \begin{split}
      \sum_{B \in \mathcal{B}}
      \Abs{\beta^{(T)}_{B,l}}
      = &
      \sum_{B \in \mathcal{B}}
      \bigg|
      \sum_{k \in B}
      \sum_{\substack{m_1, m_2 \in \mathcal{M}_{T}\\m_2 - m_1 = l}}
      \left( \eta^{(T)}_{k,m_1} \right)^{\ast}
      \eta^{(T)}_{k,m_2}
      \bigg| \\
      \leq &
      \sum_k
      \sum_{\substack{m_1, m_2 \in \mathcal{M}_{T}\\m_2 - m_1 = l}}
      \Abs{
	\eta^{(T)}_{k,m_1}
	\eta^{(T)}_{k,m_2}
      } ,
    \end{split}
  \end{equation*}
  where $\sum_k$ is the sum over all possible states $\Ket{k}$.
  From (\ref{eqn:eta-norm}) we now derive by the Cauchy-Schwarz inequality
  \begin{equation*}
    \begin{split}
      &
      \sum_k
      \sum_{\substack{m_1, m_2 \in \mathcal{M}_{T}\\m_2 - m_1 = l}}
      \Abs{
	\eta^{(T)}_{k,m_1}
	\eta^{(T)}_{k,m_2}
      }
      =
      \sum_k
      \sum_{m : m, m+l \in \mathcal{M}_{T}}
      \Abs{
	\eta^{(T)}_{k,m}
	\eta^{(T)}_{k,m+l}
      } \\
      \leq &
      \sum_k
      \left(
	\sum_{m \in \mathcal{M}_{T}}
	\Abs{\eta^{(T)}_{k,m}}^2
      \right)^{1/2}
      \left(
	\sum_{m + l \in \mathcal{M}_{T}}
	\Abs{\eta^{(T)}_{k,m+l}}^2
      \right)^{1/2}
      \leq
      \sum_k
      \sum_{m \in \mathcal{M}_{T}}
      \Abs{\eta^{(T)}_{k,m}}^2
      \leq
      1 .
    \end{split}
  \end{equation*}
  It remains to show that the two definitions of $\mathcal{L}_T$ in equations (\ref{eqn:L_T}) and (\ref{eqn:L_Tv2}) are identical.
  The proof is by induction.
  $T=0$ is trivially true.
  We use the definition (\ref{eqn:M-recursive}) of $\mathcal{M}_T$ to see that
  \begin{equation*}
    \begin{split}
      \mathcal{L}_{T+1}
      = &
      \Set{ m_1 - m_2 \, : \, m_1, m_2 \in \mathcal{M}_{T+1} } \\
      = &
      \big\{ m_1 - m_2, m_1 + p_{T+1} - m_2, m_1 - m_2 - p_{T+1}, \\
      & \ \ \ \ \ \ m_1 + p_{T+1} - m_2 - p_{T+1}\, : \, m_1, m_2 \in \mathcal{M}_{T} \big\} \\
      = &
      \Set{ l, l+p_{T+1}, l-p_{T+1} \, : \, l \in \mathcal{L}_T } ,
    \end{split}
  \end{equation*}
  which completes the proof.
\end{proof}

Note that $\Abs{\mathcal{L}_T} \leq 3^T$.
This bound is sharp, since for the choice of $p_i = 3^{i-1}$ we have $\mathcal{L}_0 = \Set{0}$, $\mathcal{L}_1 = \Set{-1, 0, 1}$, $\mathcal{L}_2 = \Set{-4, -3, -2, \ldots, 3, 4}$ and in general
\begin{equation*}
  \mathcal{L}_T = \Set{-3^T - 3^{T-1} - \ldots -1, \ldots, 3^T + 3^{T-1} + \ldots +1}.
\end{equation*}

\subsection{Fourier Analysis of Power Query Algorithms}

With Theorem \ref{theo:trig-poly-ap} and Lemma \ref{lemm:pbc} we have the tools needed to provide a lower bound for the Sturm-Liouville problem.
We are now able to apply our frequency analysis technique to this problem.

\begin{theo}\label{theo:lower-bound-cap}
  Any quantum algorithm $\mathcal{A}$ with $T$ power queries which estimates the smallest eigenvalue $\lambda(q)$ in the Sturm-Liouville eigenvalue problem for all inputs $q(x) = q \in \intervalco{0}{1}$ with precision $e(\mathcal{A}, T) \leq \eps$ and probability greater than $3/4$ has to use $T = \Omega ( \log (1/\eps) )$ power queries.
\end{theo}

Notice that a lower bound on the ``easy'' subset of constant functions $q(x)=q$ implies that the same lower bound holds for any set of inputs that includes the constant functions, hence it also holds for the class $\mathbf{Q}$.
We also would like to remark that the lower bound $T = \Omega ( \log (1/\eps) )$ does not depend on the number of discretization points $n$.

\begin{proof}
  After $T$ power queries we measure the final state and receive a state $\Ket{k}$ with probability $p_k (q)$.
  From the integer $k$ we classically compute a solution $\widetilde{\lambda}(k)$.
  A successful algorithm has to return an  $\eps$-approximation for every $q \in \intervalco{0}{1}$ with probability
  \begin{equation*}
    \sum_{ k : \Abs{ \lambda(q) - \widetilde{\lambda}(k) } \leq \eps} p_{k}(q) \geq \frac{3}{4},
  \end{equation*}
  see Definition \ref{defi:sl-power-query-algo}.
  Define
  \begin{equation*}
    A_{q, \eps} := \set{ k : \abs{ \lambda(q) - \widetilde{\lambda}(k) } \leq \eps}
  \end{equation*}
  as the set of states that are mapped to $\eps$-correct answers for input $q$.
  Choose $N \in \mathbb{N}$ such that $\frac{1}{N}$ is slightly bigger than $2 \eps$, i.e.,
  $
    \frac{1}{N+1} \leq 2 \eps < \frac{1}{N}
  $
  and define the points $x_r := (r+1/2)/N$ for $r=0,1, \ldots,N-1$.
  For the inputs $q=x_r$ we 
  can visualize the quantum algorithm $\mathcal{A}$ as in Figure \ref{fig:algomapping}.
  \begin{figure}[!tbh]
  \begin{center}
\ifx\pdfoutput\undefined%
  \includegraphics[angle=270,width=\columnwidth]{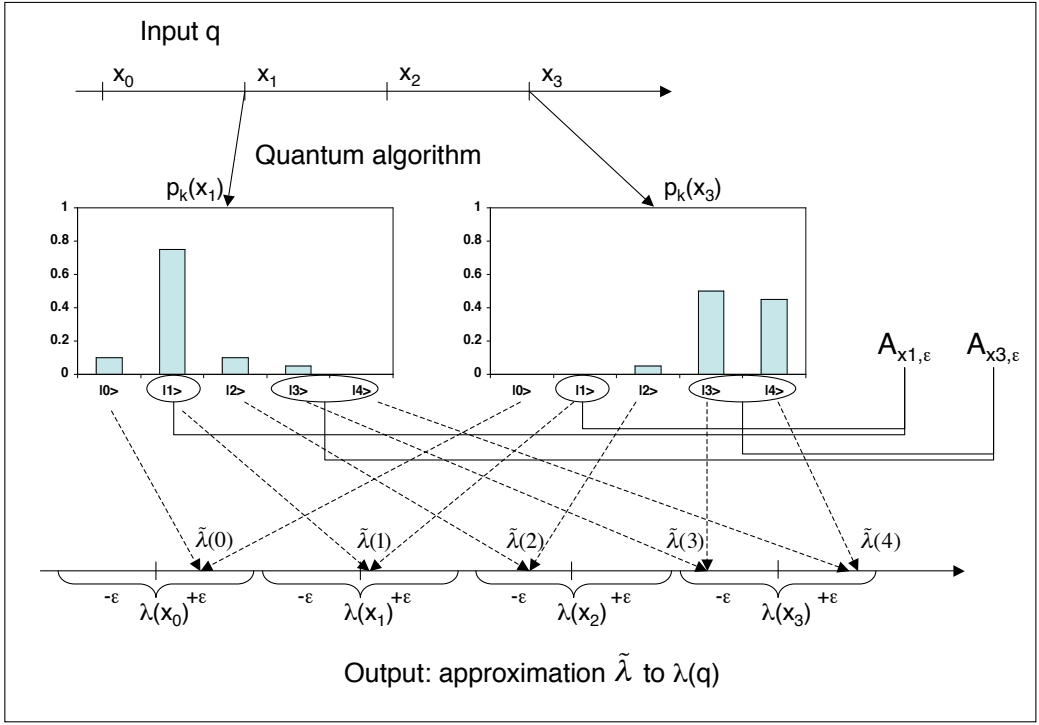}%
\else%
  \includegraphics[width=\columnwidth]{algomapping_cropped}%
\fi%
  \caption{A quantum algorithm for the Sturm-Liouville problem with inputs $q=x_r$, $r=0, \dots, N-1$, will result in a probability distribution $p_k(q)$ on the states $\Ket{k}$ that are measured.
    Each state $\Ket{k}$ is mapped to an answer $\widetilde{\lambda}(k)$.
    We write $A_{x_r,\epsilon}$ for the set of all states $\Ket{k}$ that are mapped to $\eps$-approximations of $\lambda(x_r)$.
    \label{fig:algomapping}}%
  \end{center}%
  \end{figure}
  Notice that the sets
  $A_{x_r, \eps}$
  are mutually disjoint for $r=0, \ldots, N-1$, because $x_r$ and $x_{r+1}$ are chosen such that
  \begin{equation*}
  \Abs{\lambda(x_r) - \lambda(x_{r+1})}
  =
  \Abs{
  16 \sin^2 \big( \tfrac{\pi}{4} \big) + \tfrac{r+\frac{1}{2}}{N}
  - 16 \sin^2 \big( \tfrac{\pi}{4} \big) - \tfrac{r+1+\frac{1}{2}}{N}
  }
  =
  \frac{1}{N}
  > 2 \eps .
  \end{equation*}
  Therefore there can be no state $\Ket{k}$ that is mapped to an output $\widetilde{\lambda}(k)$, which is an $\eps$-approximation to $\lambda(x_r)$ and $\lambda(x_{r+1})$ at the same time.
  Let 
  \begin{equation}
    \label{eqn:preps}
    p_{r,\eps}(q) = \sum_{k \in A_{x_r,\eps}} p_k(q)
  \end{equation}
  be the probability of measuring an $\eps$-appro\-xi\-ma\-tion to $\lambda(x_r)$.
  Since the sets $A_{x_r, \eps}$ partition the set of all outputs, Lemma \ref{lemm:pbc} allows us to write
  \begin{equation*}
    p_{r,\eps} (q)
    =
    p_{A_{x_r,\eps}} (q)
    =
    \sum_{l \in \mathcal{L}_{T}}
    \beta^{(T)}_{r,\eps,l}
    e^{ \tfrac{i}{2} l q }.
  \end{equation*}

  We apply the $N$-point inverse discrete Fourier Transform to $p_{r,\eps}(q)$, which we evaluate at the points $x_n$, and get the following value at $k = 0, 1, \ldots, N-1$:
  \begin{equation}\label{eqn:fourier-beta}
    \begin{split}
      DFT_N[p_{r,\eps}](k)
      = &
      \sum_{n=0}^{N-1}
      p_{r,\eps} (x_n)
      e^{ - 2 \pi i k n/N } \\
      = &
      \sum_{n=0}^{N-1}
      \sum_{l \in \mathcal{L}_{T}}
      \beta^{(T)}_{r,\eps,l}
      e^{ \tfrac{i}{2} l (n+1/2)/N }
      e^{ - 2 \pi i k n/N } \\
      = &
      \sum_{l \in \mathcal{L}_{T}}
      \beta^{(T)}_{r,\eps,l}
      e^{ \tfrac{i}{2} l/(2N) }
      \sum_{n=0}^{N-1}
      e^{ 2 \pi i (\frac{l}{4 \pi} - k) n/N } \\
      = &
      \sum_{l \in \mathcal{L}_{T}}
      \beta^{(T)}_{r,\eps,l}
      e^{ \tfrac{i}{2} l/(2N) }
      \left\{
      \begin{array}{ll}
	\frac{e^{2 \pi i (\frac{l}{4 \pi} - k)} - 1}{e^{2 \pi i (\frac{l}{4 \pi} - k) / N} - 1} \hspace{-3mm} & , \frac{l}{4 \pi} \not\equiv k \hspace{-3mm} \pmod{N} \\
	N & , \frac{l}{4 \pi} \equiv k\hspace{-3mm} \pmod{N}
      \end{array}
      \right\}
    \end{split}
  \end{equation}
  where $\frac{l}{4 \pi} \equiv k \pmod{N}$ indicates that there exists an integer $z$ such that $\frac{l}{4 \pi} = k + z N$.
  For every $l$ define $l_{/4 \pi (N)} \in \intervalco{0}{N}$ as
  \begin{equation*}
    l_{/4 \pi (N)} := \min \Set{ l/(4 \pi) - z N \, : \, z = 0, 1, 2, ..., \text{ and } l/(4 \pi) - z N \geq 0}.
  \end{equation*}
  Then $\exp(2 \pi i \frac{l}{4 \pi}/N) = \exp(2 \pi i l_{/4 \pi (N)} / N)$.
  To take the absolute value of equation (\ref{eqn:fourier-beta}),
  we use
  $
    \Abs{e^{i 2 \theta} - 1}
    =
    2 \Abs{\sin (\theta)}
  $
  and get
  \begin{multline}\label{eqn:fourier-beta-abs}
    \Abs{DFT_N[p_{r,\eps}](k)}
    \leq
    \sum_{l \in \mathcal{L}_{T}}
    \Abs{\beta^{(T)}_{r,\eps,l}}
    \left\{
    \begin{array}{ll}
      \frac{
	\Abs{
	  \sin (\pi (l_{/4 \pi (N)} - k))
	}
      }{
	\Abs{
	  \sin (\pi (l_{/4 \pi (N)} - k)/N)
	}
      } & , l_{/4 \pi (N)} \neq k \\
      N & , l_{/4 \pi (N)} = k
    \end{array}
    \right\}
  \end{multline}

  We can bound the Fourier transform (\ref{eqn:fourier-beta})
  by separating the correct answers, i.e., the $\eps$-approximations to $x_r$, from the rest:
  if the input $q = x_r$ then the algorithm has to return an answer $\widetilde{\lambda}$ that is $\eps$-close to the correct answer $\lambda(x_r)$ with probability greater or equal $3/4$.
  This probability is given by $p_{r,\eps} (q)$, i.e., we demand that $p_{r,\eps}(x_r) \geq 3/4$.
  Then:
  \begin{equation}\label{eqn:fourier-lower}
  \begin{split}
      \Big| \sum_{n=0}^{N-1} p_{r,\eps}( x_n ) e^{- 2 \pi i k n/N } \Big|
      \geq &
      \Big| p_{r,\eps}( x_r ) \Big|
      - \sum_{\substack{n=0\\n \neq r}}^{N-1} \Big| p_{r,\eps}( x_n ) \Big|  \\
      \geq &
      \frac{3}{4}
      - \sum_{\substack{n=0\\n \neq r}}^{N-1} p_{r,\eps}( x_n ) ,
  \end{split}
  \end{equation}
  Consider the second term in (\ref{eqn:fourier-lower}),
  $
  \sum_{\substack{n=0\\n \neq r}}^{N-1} p_{r,\eps}( x_n )
  $.
  Recall that $p_{r,\eps} (q)$ is the probability that the algorithm measures a state $\Ket{k}$ that is mapped to an answer $\widetilde{\lambda}(k)$ that is an $\eps$-approximation to $\lambda(x_r)$, i.e., $\Ket{k} \in A_{x_r,\eps}$, see (\ref{eqn:preps}).
  This probability $p_{r,\eps}(q)$ depends on the actual input $q$.
  For input $q = x_n \neq x_r$ a state $\Ket{k} \in A_{x_r,\eps}$ will \emph{not} yield an $\eps$-correct answer:
  we chose the $x_n$, $n=0, \ldots, N-1$, such that $\Abs{ \lambda(x_n) - \lambda(x_r) } > 2 \eps$ for $n \neq r$, and thus there cannot be an $\eps$ close answer for both $x_r$ and $x_n$.
  The sum
  $
  \sum_{\substack{n=0\\n \neq r}}^{N-1} p_{r,\eps}( x_n )
  $
  now tells us how often the algorithm chooses a state from $A_{r,\eps}$.

  If we knew that none of the wrong answers is preferred by our algorithm, say e.g.
  $
    \sum_{\substack{n=0\\n \neq r}}^{N-1} p_{r,\eps}( x_n ) < \frac{1}{2}
  $,
  equation (\ref{eqn:fourier-lower}) would read
  \begin{equation}\label{eqn:fourier-lower-bound}
    \Big| \sum_{n=0}^{N-1} p_{r,\eps}( x_n ) e^{- \frac{2 \pi i k n}{N} } \Big|
    \geq
    \frac{3}{4}
    - \sum_{\substack{n=0\\n \neq r}}^{N-1} p_{r,\eps}( x_n )
    >
    \frac{1}{4} .
  \end{equation}

  We will show that this property has to be true for some $r=0, \ldots, N-1$, indexing the set of states $A_{x_r,\eps}$ that represents numbers $\eps$-close to $x_r$.
  Let $R^<$ be the set of all $r$ for which
  $
    \sum_{\substack{n=0\\n \neq r}}^{N-1} p_{r,\eps}( x_n ) < \frac{1}{2}
  $
  holds and $R^\geq$ the set for which it does not.
  We estimate the number of elements of $R^<$ by splitting
  \begin{equation*}
    N
    =
    \sum_{n = 0}^{N-1}
    1
    \geq
    \sum_{n = 0}^{N-1}
    \sum_{r = 0}^{N-1}
    p_{r,\eps} (x_n)
    =
    \sum_{r = 0}^{N-1}
    p_{r,\eps} (x_r)
    +
    \sum_{r = 0}^{N-1}
    \sum_{\substack{n=0\\n \neq r}}^{N-1}
    p_{r,\eps} (x_n)
  \end{equation*}
  into the following parts:
  \begin{alignat*}{3}
    N
    & \geq
    \sum_{r = 0}^{N-1}
    p_{r,\eps} (x_r) 
    && +
    \sum_{r \in R^<}
    \sum_{\substack{n=0\\n \neq r}}^{N-1}
    p_{r,\eps} (x_n)
    && +
    \sum_{r \in R^\geq}
    \sum_{\substack{n=0\\n \neq r}}^{N-1}
    p_{r,\eps} (x_n) \\
    & \geq
    N \frac{3}{4}
    && + \Abs{R^<} \cdot 0
    && + \Abs{R^\geq} \frac{1}{2}
  \end{alignat*}
  and therefore we can conclude that
  $
    \Abs{R^\geq}
    \leq
    \frac{1}{2} N
  $ and thus
  $
    \Abs{R^<}
    \geq
    \frac{1}{2} N
  $.
  Now $\Abs{R^<} > 0$ implies that we can actually choose an element $r \in R^<$.
  Fix such an $r$ and we can combine equations (\ref{eqn:fourier-beta-abs}) and (\ref{eqn:fourier-lower-bound}) to
  \begin{equation}\label{eqn:fourier-final}
    1/4
    <
    \sum_{l \in \mathcal{L}_{T}}
    \Abs{\beta^{(T)}_{r,\eps,l}}
    \left\{
    \begin{array}{ll}
      \frac{
	\Abs{
	  \sin (\pi (l_{/4 \pi (N)} - k))
	}
      }{
	\Abs{
	  \sin (\pi (l_{/4 \pi (N)} - k)/N)
	}
      } & , l_{/4 \pi (N)} \neq k \\
      N & , l_{/4 \pi (N)} = k
    \end{array}
    \right\} .
  \end{equation}
  We will now fix the parameter $k=0,1,\ldots,N-1$ in inequality (\ref{eqn:fourier-final}) in such a way that the terms in the sum on the right-hand-side of the inequality are as small as possible.
  This will imply that the sum must be over a large number of elements, i.e., that $\Abs{\mathcal{L}_T}$ is large.
  Since $\Abs{\mathcal{L}_T} \leq 3^T$ this will help us to ultimately prove that $T = \Omega( \log N)$ if we could show that there is an $\alpha > 0$ such that $\Abs{\mathcal{L}_T}^{\alpha}  = \Omega ( N )$.
  More specifically we will show that $\Abs{\mathcal{L}_T}^2 \geq \frac{1}{10} N$ which proves $T = \Omega( \log N)$.
  
  We prove $\Abs{\mathcal{L}_T}^2 \geq \frac{1}{10} N$ by contradiction.
  Assume $\Abs{\mathcal{L}_T}^2 < \frac{1}{10} N$.
  This assumption allows us to find a $k$ such that the right-hand-side of inequality (\ref{eqn:fourier-final}) is smaller than the left-hand-side, which will lead to our desired contradiction.

  If we project $\mathcal{L}_T$ into the interval $\intervalco{0}{N}$ through $l \mapsto l_{/4 \pi (N)}$ we will get a set $\Set{l_{/4 \pi (N)} \, : \, l \in \mathcal{L}_T}$.
  Order this set as $0 \leq t_1 \leq t_2 \leq \ldots \leq t_{\Abs{\mathcal{L}_T}} < N$.
  This defines ``gaps'' between these numbers, i.e., intervals $G=\intervaloo{t_j}{t_{j+1}}$ for $j=1, \ldots, \Abs{\mathcal{L}_T}$ if we define $t_{\Abs{\mathcal{L}_T}+1} = t_1+N$ (we ``wrap around'').
  Define the width $w(G)$ of such a gap $G$ as the distance between its endpoints.
  Thus $w(\intervaloo{t_j}{t_{j+1}}) = t_{j+1}-t_j$.
  
  Let $G_m$ be the gap with the maximal width $w(G_m)$ in the distribution.
  Its width must be
  $w(G_m) \geq N / \Abs{\mathcal{L}_T}$, since
  \begin{equation*}
    N = \sum_G w(G)
    \leq \sum_G \max_G w(G)
    = \Abs{\mathcal{L}_T} \max_G w(G).
  \end{equation*}
  Additionally $w(G_m) > 10$, since we assumed $\Abs{\mathcal{L}_T}^2 < \frac{1}{10} N$ and therefore
  $
    \frac{N}{\Abs{\mathcal{L}_T}} > 10 \Abs{\mathcal{L}_T} \geq 10.
  $
  Thus there are at least ten integers $k \in \Set{0, 1, \ldots, N-1}$ that fall into this largest gap $G_m$, i.e $k \in G_m$.
  One of these $k$ has maximum distance to \emph{both} boundaries $t_j$ and $t_{j+1}$ of $G_m$: it is the $k$ that is closest to the middle $m = \frac{t_{j+1}+t_j}{2}$ of $G_m = \intervaloo{t_j}{t_{j+1}}$.
  This integer $k$ fulfills $\Abs{k-m} \leq \frac{1}{2}$ and
  \begin{equation*}
    \begin{split}
      \min \Set{ k-t_j, t_{j+1}-k}
      & =
      \min \Set{ m-t_j + k-m, m-k + t_{j+1}-m} \\
      & \geq
      \frac{w(G_m)}{2} - \frac{1}{2}
      \geq
      \frac{N}{2 \Abs{\mathcal{L}_T}} - \frac{1}{2}.
    \end{split}
  \end{equation*}
  Fix this $k \in \intervaloo{t_j}{t_{j+1}}$.
  Now $\Abs{\sin(x)} \geq 2/\pi \Abs{x}$ for $-\pi/2 \leq x \leq \pi/2$ and therefore
  \begin{equation*}
    \begin{split}
      \min_{l \in \mathcal{L}_T}
      \Abs{\sin \frac{\pi (l_{/4 \pi (N)} - k)}{N}}
      \geq &
      \min_{l \in \mathcal{L}_T}
      \frac{2}{N} \Abs{l_{/4 \pi (N)} - k}
      =
      \frac{2}{N} \min \Set{ k-t_j, t_{j+1}-k} \\
      \geq &
      \frac{1}{\Abs{\mathcal{L}_T}} - \frac{1}{N}.
    \end{split}
  \end{equation*}
  Then we can use this to estimate (\ref{eqn:fourier-final}):
  \begin{equation*}
    1/4
    <
    \sum_{l \in \mathcal{L}_{T}}
    \Abs{\beta^{(T)}_{r,\eps,l}}
    \frac{
      \Abs{
        \sin (\pi (l_{/4 \pi (N)} - k))
      }
    }{
      \Abs{
        \sin (\pi (l_{/4 \pi (N)} - k)/N)
      }
    }
    \leq
    \sum_{l \in \mathcal{L}_{T}}
    \Abs{\beta^{(T)}_{r,\eps,l}}
    \frac{1}{1/\Abs{\mathcal{L}_T} - 1/N}
  \end{equation*}
  We sum the last inequality over all $r \in R^<$ for which it is valid, and get:
  \begin{equation}\label{eqn:last-inequality}
    \sum_{r \in R^<}
    \frac{1}{4}
    \leq
    \frac{1}{1/\Abs{\mathcal{L}_T} - 1/N}
    \sum_{r \in R^<}
    \sum_{l \in \mathcal{L}_{T}}
    \Abs{\beta^{(T)}_{r,\eps,l}}
  \end{equation}
  Since the number of elements in $R^<$ is bounded by $\Abs{R^<} \geq \frac{1}{2} N$, the left-hand-side of (\ref{eqn:last-inequality}) is bounded by $\frac{1}{8} N \leq \Abs{R^<} \frac{1}{4} $.
  The right-hand-side of inequality (\ref{eqn:last-inequality}) can be bounded through Lemma \ref{lemm:pbc}:
  \begin{equation*}
    \sum_{r \in R^<}
    \sum_{l \in \mathcal{L}_{T}}
    \Abs{\beta^{(T)}_{r,\eps,l}}
    \leq
    \Abs{\mathcal{L}_T} .
  \end{equation*}
  If we put both sides together again and recall that we assumed $\Abs{\mathcal{L}_T}^2 < \frac{1}{10} N$ we get
  \begin{equation*}
    \frac{1}{8} N
    \leq
    \frac{\Abs{\mathcal{L}_T}}{1/\Abs{\mathcal{L}_T} - 1/N}
    =
    \frac{\Abs{\mathcal{L}_T}^2}{1 - \Abs{\mathcal{L}_T} / N}
    <
    \frac{\frac{1}{10} N}{1 - \frac{1}{10 \Abs{\mathcal{L}_T}}}
    \leq
    \frac{\frac{1}{10} N}{1 - \frac{1}{10}}
    =
    \frac{1}{9} N,
  \end{equation*}
  which is a contradiction.

  Therefore $\Abs{\mathcal{L}_T}^2 \geq \frac{1}{10} N$ must hold.
  This, together with $\Abs{\mathcal{L}_T} \leq 3^T$, leads us to
  $
  N \leq 10 \cdot 9^T
  $.
  Take the logarithm and we get $T = \Omega(\log N)$.
  We chose $N$ such that
  $
  \frac{1}{N+1} \leq 2 \eps < \frac{1}{N}
  $
  which finally proves that the number of power queries $T$ for any algorithm $\mathcal{A}$ with error $e(\mathcal{A},T) \leq \eps$ has to be of the order $T = \Omega(\log (1/\eps) )$.
\end{proof}

\section{Discussion}

In this paper we have proven lower bounds for the number of quantum power queries for the Sturm-Liouville problem and settled an open problem in \cite{pap-woz-05}.

How does this number of $T=\Theta(\log(1/\eps))$ power queries relate to the cost of quantum algorithms?
Here we understand ``cost'' as an abstraction on the number of elementary quantum gates or the duration for which a Hamiltonian has to be applied to a quantum system.
Suppose the function $q$ is from a class $\mathbf{Q}' \subseteq \mathbf{Q}$ where each power query $W_l^p ( \exp(\tfrac{i}{2} M_q) )$ can be implemented with $\text{cost}(W_l^p ( \exp(\tfrac{i}{2} M_q) )) = \text{cost}(\mathbf{Q}',p)$.

If we implement $W_l^p ( \exp(\tfrac{i}{2} M_q) )$ naively as
$$W_l^p ( \exp(\tfrac{i}{2} M_q) ) = \left( W_l^1 ( \exp(\tfrac{i}{2} M_q) ) \right)^p, $$
then $\text{cost}(\mathbf{Q}',p) = p \cdot \text{cost}(\mathbf{Q}')$ and the cost of the Sturm-Liouville algorithm with $T = \Theta(\log(1/\eps))$ power queries grows as
\begin{equation*}
  \sum_{j=0}^{T-1}
  \text{cost}(\mathbf{Q}', 2^j)
  =
  \sum_{j=0}^{T-1}
  2^j \cdot \text{cost}(\mathbf{Q}')
  =
  (2^T - 1) \cdot \text{cost}(\mathbf{Q}')
  =
  \Theta ( 1/\eps ) \cdot \text{cost}(\mathbf{Q}') .
\end{equation*}
This is polynomial in $1/\eps$ just like the Sturm-Liouville algorithm with bit queries discussed in \cite{pap-woz-05}.
To take advantage of the proposed power query algorithm it is therefore necessary to realize power queries $W_l^p ( \exp(\tfrac{i}{2} M_q) )$ on a quantum computer in such a way that $\text{cost}(\mathbf{Q}',p) = o ( p ) \cdot \text{cost}(\mathbf{Q}') $

The implementation of power queries with cost that is not linear in the power $p$ of the query is still not settled and requires more work.
It would be of interest to identify subclasses $\mathbf{Q}' \subseteq \mathbf{Q}$ for which we are able to prove $\text{cost}(\mathbf{Q}',p) = o ( p ) \cdot \text{cost}(\mathbf{Q}') $.

Another open question is whether it is possible to extend the methods we used for upper and lower bounds for the Sturm-Liouville problem in one dimension to similar problems in higher dimensions.
Most important for this problem is probably the extension of the results in \cite{jak-pap-03} on approximations of the eigenvector with the smallest eigenvalue to higher dimensions.

\section{Acknowledgments}
The author would like to thank M. Kwas, A. Papageorgiou, J. Traub and H. Wo{\'z}niakowski for inspiring discussions.
Special thanks to an anonymous referee, who pointed out a gap in a previous version of this paper.
Partial funding was provided by Columbia University through a Presidential Fellowship.
This research was supported in part by the National Science Foundation and the Defense Advanced Research Projects Agency.

\bibliographystyle{plain}
\bibliography{qc}

\end{document}